\documentclass[electronic]{vgtc}             

\graphicspath{{figures/}{pictures/}{images/}{./}} 

\usepackage{times}                     

\usepackage{mathptmx}                  

\onlineid{0}

\vgtccategory{MR}

\vgtcinsertpkg

\title{Conversational Interfaces for Parametric Conceptual Architectural Design: Integrating Mixed Reality with LLM-driven Interaction}

\author{
  Ruochen Ji\\
  \texttt{jirc22@mails.tsinghua.edu.cn}
  \and
  Lyu Tiangang\\
  \texttt{ltg24@mails.tsinghua.edu.cn}
}

\teaser{
  \centering
  \includegraphics[width=\linewidth]{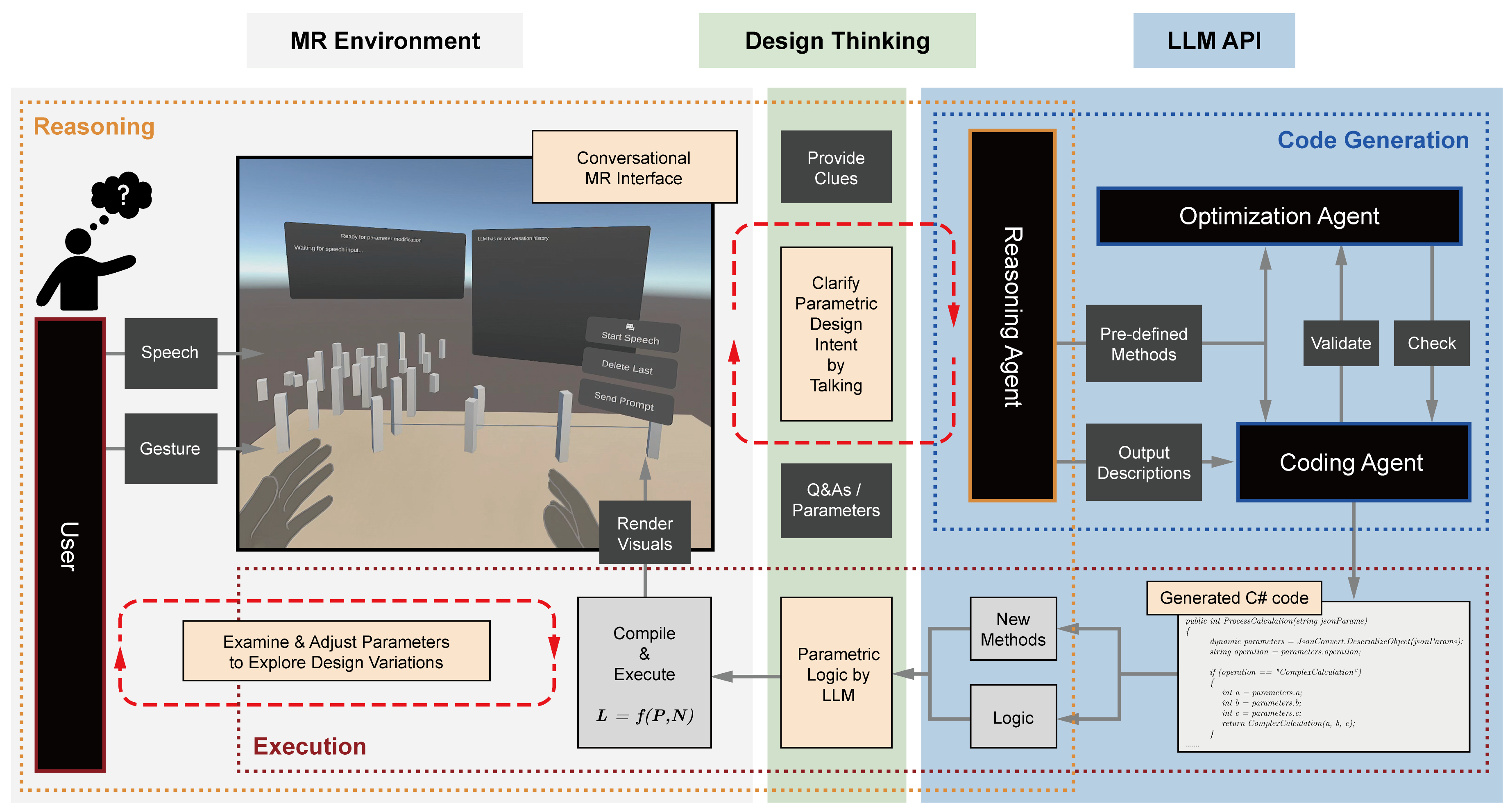}
  \caption{Overview of system design objectives. This framework establishes a comprehensive conversational parametric design interaction through a ``Reasoning"-``Code Generation"-``Execution" cycle. Users in MR may have either well-defined or ambiguous design intentions. Through talking with a Reasoning Agent LLM, design intentions are ultimately grounded in logic and parameters, then transformed by Coding/Optimization Agents into directly compilable and executable code blocks, which are returned to the MR environment for visualization. Users can explore design variations through parameter adjustment, immersive observation within the MR environment, or direct manipulation of graphical elements using gesture interactions.}
  \label{fig:overview}
}

\abstract{Mixed reality (MR) environments offer embodied spatial interaction, providing intuitive 3D manipulation capabilities that enhance the conceptual design process. Parametric modeling, a powerful and advanced architectural design method, enables the generation of complex, optimized geometries. However, its integration into MR environments remains limited due to precision constraints and unsuitable input modalities. Existing MR tools prioritize spatial interaction but lack the control and expressiveness required for parametric workflows, particularly for designers without formal programming backgrounds. We address this gap by introducing a novel conversational MR interface that combines speech input, gesture recognition, and a multi-agent large language model (LLM) system to support intuitive parametric modeling. Our system dynamically manages parameter states, resolves ambiguous commands through conversation and contextual prompting, and enables real-time model manipulation within immersive environments. We demonstrate how this approach reduces cognitive and operational barriers in early-stage design tasks, allowing users to refine and explore their design space. This work expands the role of MR to a generative design platform, supporting programmatic thinking in design tasks through natural, embodied interaction.} 


\begin{document}




\firstsection{Introduction}

\maketitle

In the context of architectural design, parametric design is an approach that describes a design symbolically based on the use of parameters, leveraging computational methods to generate optimized geometries through parameter-based symbolic representation \cite{caetano_computational_2020}. While this approach enhances creativity and efficiency in the design process \cite{gurel_cognitive_2023, lee_creativity_2015}, current parametric modeling tools such as Grasshopper\cite{noauthor_httpswwwgrasshopper3dcom_nodate} and Python present significant usability barriers. These tools predominantly rely on desktop-based screen-mouse (WIMP) interactions and require programming expertise, limiting accessibility and hindering rapid iteration between different parametric models during early design stages.

Mixed reality (MR) technologies offer embodied spatial interaction with intuitive 3D manipulation capabilities, yet integrating parametric modeling into MR environments remains challenging. As Wortmann et al. \cite{wortmann_differentiating_2017} denotes, ``Most designers already think programmatically, but having neither the time nor the inclination to learn programming skills, do not have the means to express or explore these patterns of thought.'' Previous studies confirm MR's potential for embodied conceptual design while highlighting a critical gap: the absence of an effective multimodal interface for handling precise parametric inputs and communication of complex design intentions within spatial environments.

The disconnect between immersive spatial interaction and algorithmic control represents a significant barrier to unlocking MR's full potential in parametric design workflows. Given the precision constraints of spatial interaction and complexity of parametric inputs, conversational interfaces using voice and gesture interaction emerge as a compelling alternative. Recent advances in large language models (LLMs) and multimodal interfaces present an opportunity to bridge this gap by enabling natural language understanding of design intent while maintaining the algorithmic precision required for parametric modeling.

In this paper, we present a novel conversational mixed reality interface specifically designed to address these challenges by integrating speech, gesture recognition, and LLM-based conversational AI (See \cref{fig:overview}). This work draws inspiration from recent advances in employing LLMs for robotics and 3D understanding tasks \cite{desolda_digital_2023,  friedrich_combining_2021, williams_understanding_2020, zhou_eliciting_2022}, utilizing a multi-LLM agent system as the backend for embodied interactions in MR. We applied the ``reasoning-to-code-generation" approach \cite{liang_code_2023} to the parametric modeling interface, facilitating intuitive workflows that handle ambiguous inputs, enable command correction, and dynamically manage uncertain parameters inherent in complex design tasks.

By leveraging MR's immersive capabilities to provide direct contextual prompting, the system reduces the cognitive and operational barriers traditionally associated with parametric modeling in immersive settings, transforming MR from a primarily visualization-centric role into an embodied tool for conceptual design ideation and iterative exploration.

We evaluate the system through a user study to collect data and feedback from professionals and students, followed by quantitative analysis of the collected design tasks as testing datasets. Our contributions are as follows:

\begin{enumerate}
    \item A novel conversational mixed reality interface for parametric modeling that integrates speech, gesture recognition, and multi-LLM systems.
    \item A reasoning-to-code-generation prompting framework that translates natural language and spatial references into parametric modeling operations.
    \item Empirical findings on the effectiveness of multimodal interaction for parametric design in MR environments.
\end{enumerate}

The rest of the paper consists of the following: Section 2 reviews related work in parametric modeling interfaces, conversational AI in spatial computing, MR for design tasks. Section 3 defines the problem space and design tasks addressed by our system. Section 4 details our system implementation and technical approach. Section 5 presents our experimental methodology and results from user studies. Section 6 discusses implications, limitations and future directions of our findings, and Section 7 concludes this paper.


\section{Related Works}
The research background of this paper integrates various fields, including parametric architecture design, conversational interfaces, and the integration of these domains within MR environments.

This paper's reasoning-to-code-generation approach draws inspiration from multiple related works, including \cite{ahn_as_2022, gao_pal_2022, hong_3d-llm_2023, huang_instruct2act_2023, liang_code_2023, lin_text2motion_2023,  vemprala_chatgpt_2024}. Collectively, these studies highlight the feasibility of addressing complex task-planning problems through natural language input, which aligns effectively with multimodal interaction capabilities in MR environment.

\subsection{Parametric Design in Architecture}
Caetano et al. \cite{caetano_computational_2020, gurel_cognitive_2023} systematically analyzed and categorized taxonomies of parametric and generative design logic. These frameworks significantly inform the task definition that the LLM-based system proposed in this paper addresses.

Previous studies have identified distinct advantages of parametric modeling over conventional design methods. Specifically, \cite{gurel_cognitive_2023} and \cite{lee_creativity_2015} examined the cognitive load and creativity among designers using different modeling paradigms. They highlight that parametric modeling generates a broader range of design variations and creative alternatives. Furthermore, studies such as \cite{alhazzaa_integrating_2023} and \cite{bhooshan_parametric_2017} emphasize the architectural specificity and educational value inherent in parametric modeling.

\subsection{Conversational Interfaces for 3D-Related Tasks}
Early multimodal interfaces, such as Bolt's \cite{bolt_put-that-there_1980}, highlight the foundational vision of combining gesture, voice, and spatial computing, an ambition that now becomes feasible with the advancement of LLMs within MR contexts.

Recent advances in LLM-powered agent frameworks have enabled natural language-controlled task planning and execution, prompting studies to integrate language models with 3D and MR environments. For example, \cite{buyruk_interactive_2022} and \cite{castelo-branco_algorithmic_2022} present implementations that link algorithmic or robotic fabrication systems with virtual reality interfaces, offering immersive spatial design agency to users.

Meanwhile, \cite{gao_pal_2022} and \cite{liang_code_2023} propose language-as-policy pipelines and program-aided prompting, establishing architectures where natural language directly informs computational behavior. \cite{cheng_empowering_2024} and \cite{hong_3d-llm_2023} further demonstrate how affordance prompting and 3D world embedding enhance LLM reasoning in spatial contexts. \cite{liu_llmp_2023} investigates the optimal planning capabilities of LLMs. Similarly, \cite{alrashedy_generating_2024} proposed pipelines that translate textual and visual prompts into CAD-like outputs. Other studies, including \cite{cai_3description_2024, wang_describe_2023, yin_text2vrscene_2024} have introduced frameworks for interactive 3D modeling using conversational inputs, emphasizing multi-turn planning, feedback loops, and scene reasoning. Additionally, \cite{sun_3d-gpt_2023} and \cite{weng_dream_2024} demonstrate procedural 3D modeling feasibility via LLMs but expose challenges such as fidelity, prompt ambiguity, and parametric controllability. These studies collectively raise foundational questions regarding task segmentation, retrieval, and execution within spatial logic systems.

Generative AI approaches also contribute significantly, with frameworks such as \cite{hollein_text2room_2023} and \cite{weng_dream_2024} that bridge speech and text-to-3D modeling through generative pipelines. Evaluative studies on conversational interfaces, such as \cite{baig_analyzing_2018}, suggest that novice users often feel more comfortable using multimodal interfaces compared to expert users, indicating that conversational interfaces could mitigate knowledge barriers inherent in current tools.

\subsection{Complex Modeling Within MR Environments}
Current studies frequently explore MR's embodied interaction capabilities, particularly in basic 3D modeling or sculpting contexts. \cite{chatterjee_free-form_2024} provides an extensive review of free-form shape modeling in XR, highlighting input, output, and workflow considerations highly relevant to parametric MR systems. Continuing studies delve into integrating multimodal interfaces with spatial modeling systems, exemplified by \cite{desolda_digital_2023,  friedrich_combining_2021, williams_understanding_2020, zhou_eliciting_2022}, exploring speech and gesture combinations for object manipulation within MR and AR contexts. Additionally, \cite{baig_analyzing_2018} and \cite{giunchi_mixing_2021} address interaction challenges encountered when integrating gesture and speech within VR design tools.

Although existing research thoroughly discusses interactions with MR environments, introducing complex modeling tasks raises unanswered questions, particularly around the precision and logical structuring necessary for parametric designs and the lack of suitable interaction paradigms specifically designed for MR.

Prior to the rise of LLMs, Alibay et al.\cite{alibay_usability_2017} highlighted challenges executing precision tasks using multimodal interactions. Similarly, Fang \& Kao et al. \cite{fang_comparisons_2023} evaluated differences in user experience between parametric CAD and free-form VR, concluding that VR enhances creative exploration while traditional CAD remains superior for precision-oriented tasks. Studies by Coppens et al. \cite{coppens_parametric_2018}, Salim et al. \cite{salim_system_2010}, Alhazzaa et al. \cite{alhazzaa_integrating_2023}, and Buyruk \cite{buyruk_interactive_2022} reveal the potential of immersive environments for intuitive spatial reasoning and manipulation. Yet, these studies show that MR primarily serves as a visualization medium, whereas the underlying logic and parameters continue to depend on conventional tools such as Grasshopper.


\section{Design Objective}

Architectural conceptual design fundamentally differs from direct generative tasks like text-to-image synthesis, as it requires iterative exploration of geometric variations and combinations to progressively align with designers' intents. Although early-stage parametric geometries typically lack detailed textures or realism, they carry meaningful implications for practical engineering and design feasibility. Thus, determining how these primitives are combined, and efficiently exploring their possibilities, is critical to the conceptual design workflow \cite{caetano_computational_2020}.

Parametric modeling logic often contains uncertainty in its types of parameters, and processing methods. While one might want to define LLM tasks as action planning using existing skills, an idea similar to automating visual programming in Grasshopper, previous research shows that mapping from language instructions to code snippets is a more versatile approach \cite{liang_code_2023}.

\begin{figure}[tb]
 \centering 
 \includegraphics[width=\columnwidth]{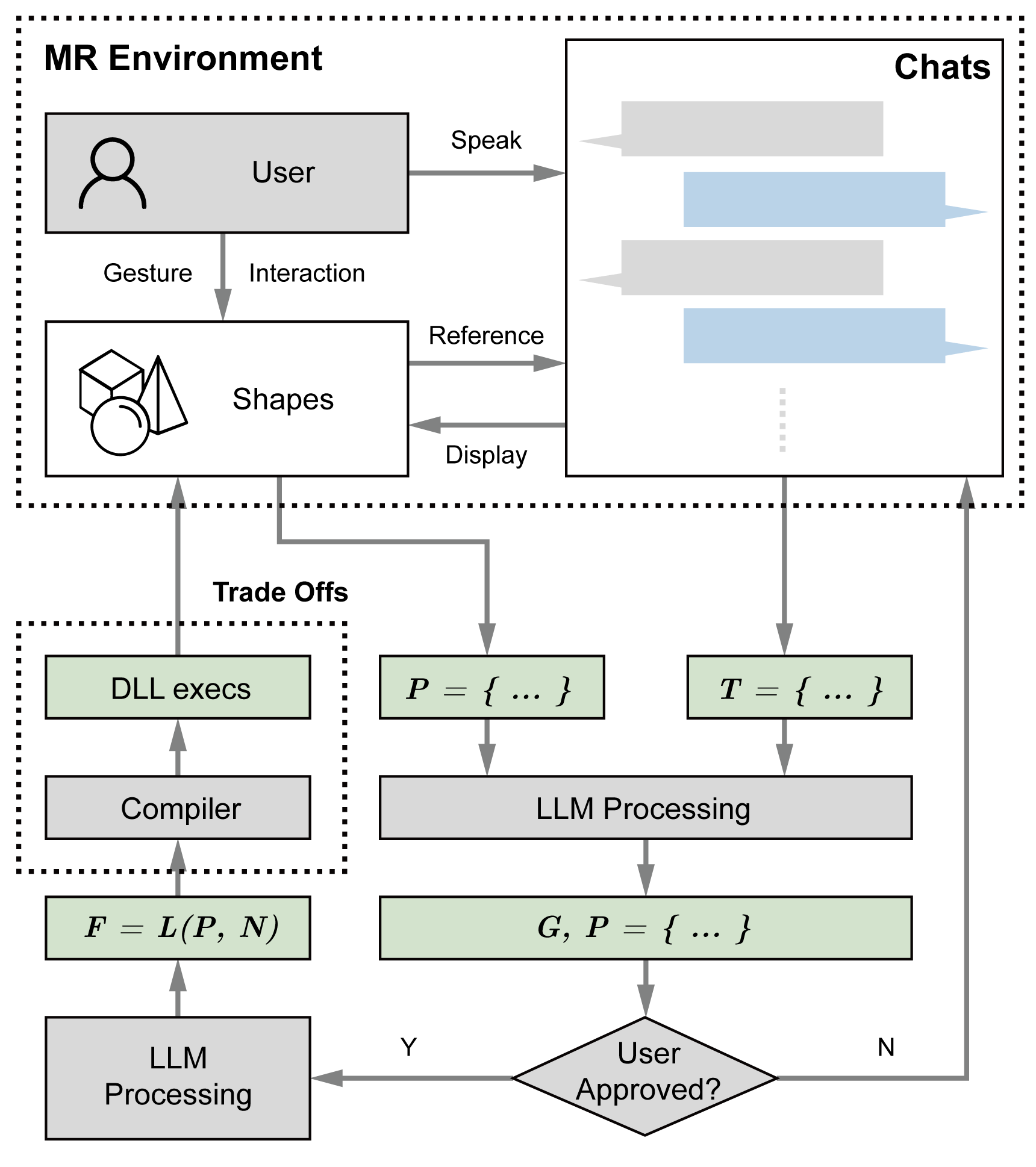}
 \caption{System Data Flow}
 \label{fig:data_flow}
\end{figure}

\subsection{Task Definition}

We propose a novel system operating within MR environment that processes user-provided \textbf{textual instructions}, denoted as $T$, to determine \textbf{design intentions} $G$ and \textbf{parameter sets} $P$ through interactive dialogue. The system ultimately produces a \textbf{parametric design function} $F$, whose execution results are visualized within MR environment (See \cref{fig:data_flow}). Users can iteratively modify parameters $P$ through gesture interactions to compare different outcomes or continue providing textual instructions. Specifically:

\begin{itemize}
    \item \textbf{Textual Instructions} $T = \{T_1, T_2, \cdots, T_N\}$ comprises a collection of $N$ dialogue turns, containing contextual information from the MR environment.
    \item \textbf{Design Intentions} $G$ documents the relationships between parameters in $P$ and encapsulates the generative logic of visual elements.
    \item \textbf{Parameter Sets} $P = \{P_1, P_2, \cdots, P_N\}$ consists of both non\-graphical and graphical parameters, with predefined selectable types.
    \item \textbf{Parametric Design Function} $F = f(P)$ represents a compilable and executable code segment for MR devices, which may include new methods and invoke existing graphical APIs.
\end{itemize}

\subsection{Implementation Trade-offs}

The hardware platform imposes runtime compilation constraints that prevent direct execution of string-based code. Instead, compilation to DLL format is required prior to execution, and linking errors occur when $F$ includes newly defined methods. To address these limitations, we change $F$'s compilation and execution into following approach (See \cref{fig:data_flow}):

\begin{enumerate}
    \item Isolate the \textbf{parametric design logic} $L(P, N)$ from $F$ as a separate file and extract the set of \textbf{newly defined methods} $N$.
    \item $N=\{N_1, \cdots\}$ are compiled into DLL files on a locally-connected x64 architecture PC as external compiler.
    \item After successful compilation of $N$ (where $N$ may be empty), $L$ is compiled, enabling the system to invoke the generated DLL files.
\end{enumerate}

\subsection{Multi-LLM Agents}

To achieve the system objectives, we implement a collaborative framework of three specialized LLM Agents:

\begin{itemize}
    \item \textbf{Reasoning Agent} $R_A$: Engages directly with users through dialogue, analyzes user intentions and parametric logic constructions, manages UI elements, and produces task descriptions for subsequent agents.
    \item \textbf{Coding Agent} $C_A$: Generates $F = L(P, N)$ and other data useful for UI controls based on specifications provided by $R_A$.
    \item \textbf{Optimization Agent} $O_A$: Performs syntax validation on $F$, cross-references with predefined libraries and previously generated method DLLs to prevent compilation errors.
\end{itemize}

Detailed prompt design is presented in the section ``Agent Prompts".


\section{Method}

Our proposed system consists of a Meta Quest 3 head-mounted display and an x64 Windows host machine. Applications deployed on both devices are developed using Unity. The host and the HMD communicate via a local area network, wherein the host functions exclusively to address the technical limitation of the Quest 3, specifically its inability to perform runtime code compilation. Consequently, the host serves as a compilation client, while the Quest 3 operates as a server.

\cref{fig:overview} illustrates the overall design objectives of the system. Data flows across two execution domains within the system: the MR environment powered by Quest 3 hardware, and cloud services that leverage existing LLM APIs. The implementation method detailed in this section supports two major iterative interactions for users. Firstly, a conversational interface guides users systematically from ambiguous initial design concepts to clearly articulated, logically sound, and parameter-referenced parametric design specifications. Secondly, once parametric design output is executed and visualized within the MR environment, users can explore the parametric design space in an embodied experience, comparing an extensive range of design alternatives by dynamically adjusting both graphical and nongraphical parameters.

\begin{figure}[tb]
 \centering 
 \includegraphics[width=\columnwidth]{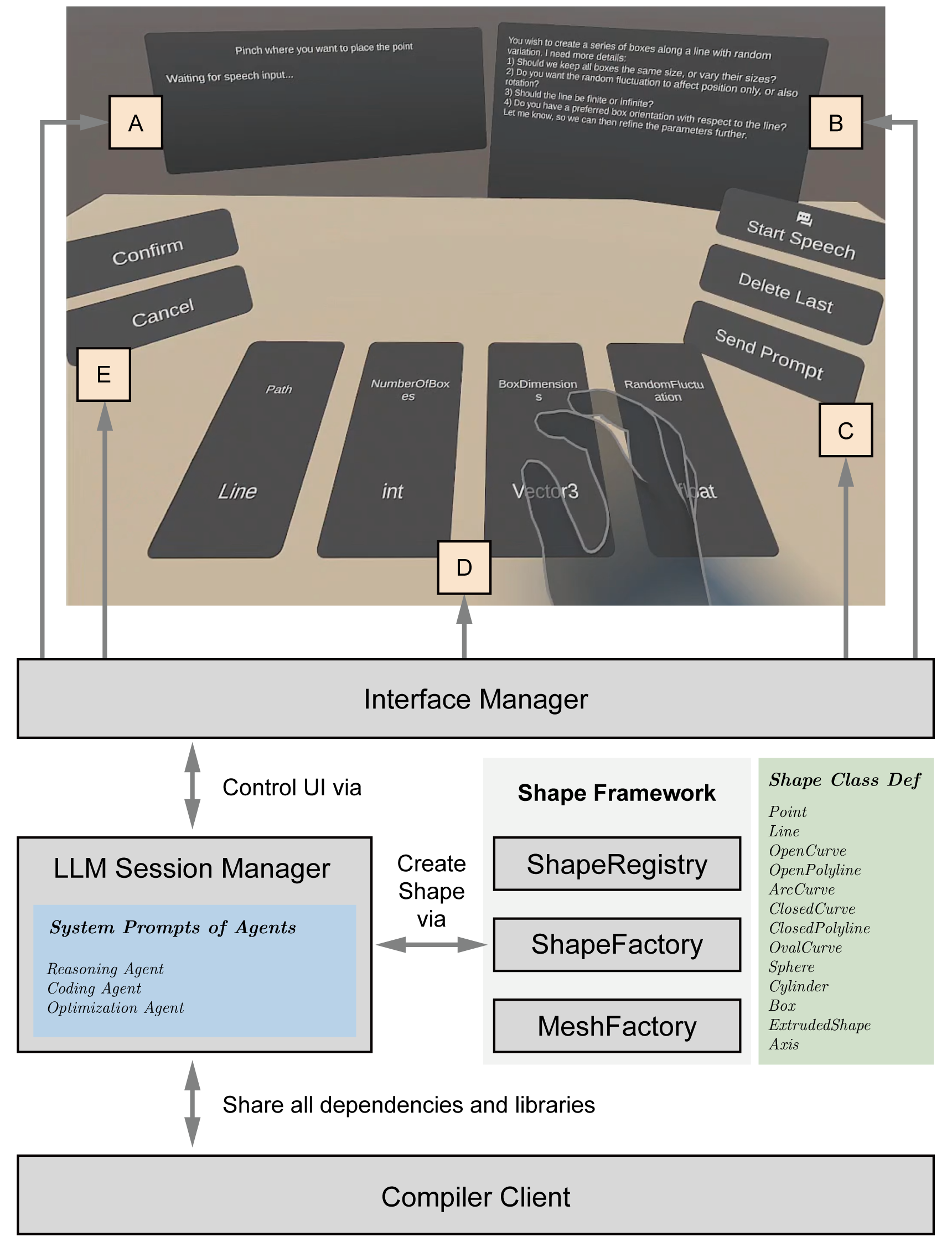}
 \caption{Overview of system modules and user interface}
 \label{fig:ui}
\end{figure}

\begin{figure}[tb]
 \centering 
 \includegraphics[width=\columnwidth]{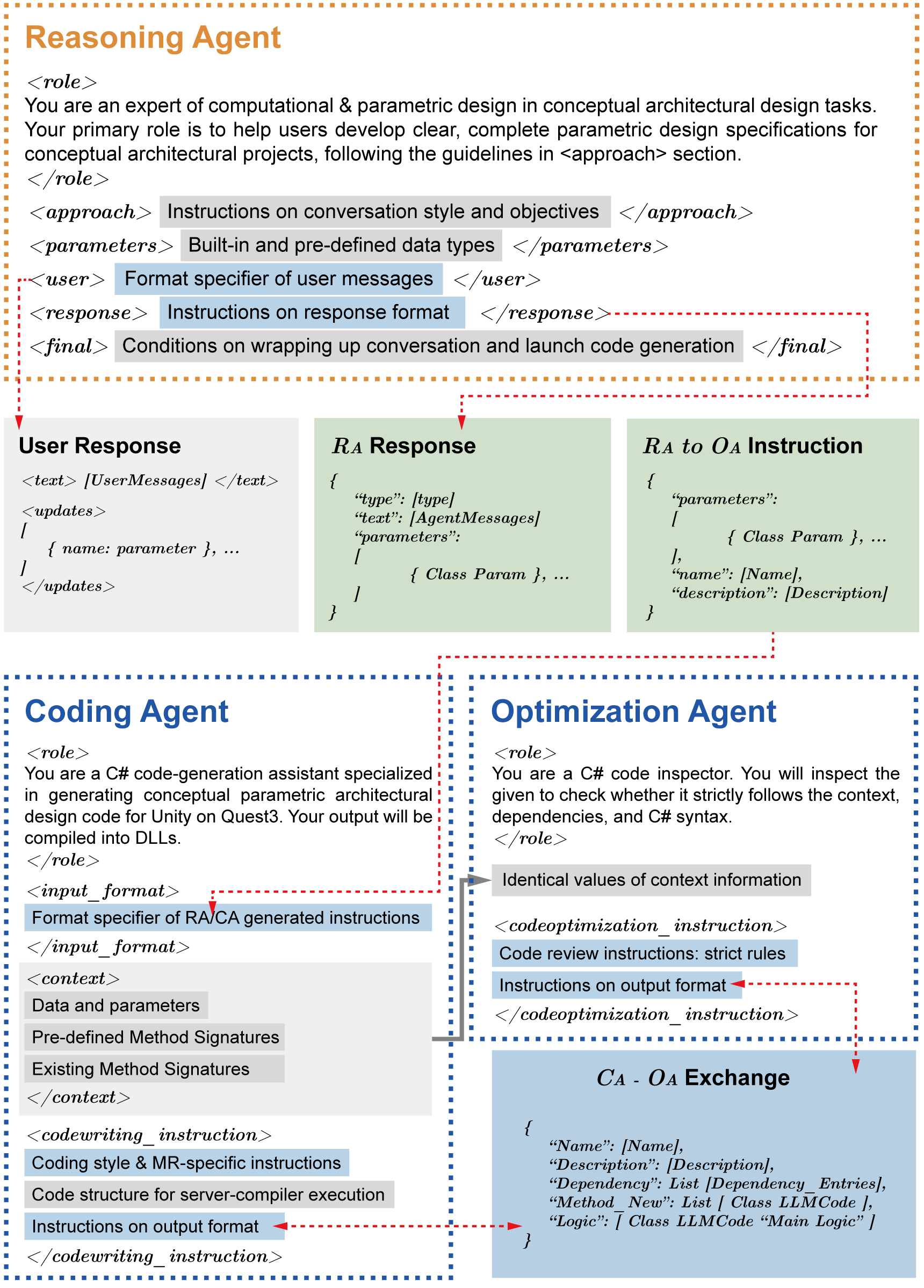}
 \caption{Prompt structure and data exchange format of agents}
 \label{fig:prompt}
\end{figure}

\subsection{User Interface and Shape Framework}

\cref{fig:ui} illustrates the four core modules that comprise our conversational interfaces:

\begin{itemize}
    \item \textbf{Interface Manager}: Coordinates UI event handling, processes audio inputs, and manages parameter adjustment interfaces. This module acts as the primary mediator between user interactions and the underlying UI components and scene objects.
    \item \textbf{ShapeFramework}: Comprises three interdependent subclasses: the $ShapeRegistry$ class, which maintains a repository of shapes and supports efficient selection and modification; the $ShapeFactory$ class, which instantiates parameterized shape objects based on user and $R_A$ specifications; and the $MeshFactory$ class, which generates visual mesh representations.
    \item \textbf{Compiler Client}: Serves as an intermediary application layer that manages code compilation processes, converting $C_A$ and $O_A$ generated code into executable files.
    \item \textbf{LLM Session Manager}: Functions as the central orchestration module for routing system-wide interactions. Through structured prompt engineering, three agents produce responses consistently formatted as JSON objects. These standardized responses act as a bidirectional data interface protocol, enabling formalized communication between agents and other core modules. This effectively embeds information into text-based conversations bridging user input modalities and LLM-generated outputs.
\end{itemize}

\subsubsection{User Interface}

During system interaction, users operate within a 3D modeling environment that simultaneously displays virtual objects and real-world surfaces, such as tables, via video pass-through. Immersive mode is enabled in the figures. Users primarily utilize gestural inputs for geometric manipulation and menu navigation.

The interface components illustrated in \cref{fig:ui} include:

A: State Window, which provides contextual interaction guidance and displays real-time dictation of user voice inputs.

B: LLM Response Window, which presents feedback message from the $R_A$.

C: Voice Input Control Panel, which manages speech recognition functionality.

D: Parameter Window, a dynamically configurable interface co-managed by the LLM and user. $R_A$ adjusts parameter types and numbers based on dialogue progression, while users specify non-geometric parameters or create geometric references by hand gesture.

E: Confirmation Panel, which appears during geometric editing operations or parameter specification tasks to validate user intent.

\subsubsection{Shape Type Framework}

We employed a 3D shape framework for parametric design, including points, curves, surfaces, and basic 3D forms. The framework defines meaningful inheritance relationships using abstract classes and interfaces, and distinct transformations between certain shape classes.

As illustrated in \cref{fig:ui}, LLM Session Manager executes parametric operations by invoking shape instantiation commands from $ShapeFactory$, as well as pre-defined manipulation methods embedded within each geometric class. Complex commands generated by the $C_A$ are composed of these foundational operations defined in the $ShapeFramework$.

This extensible framework supports advanced spatial operations, such as random distribution, sweeping, and lofting, without specialized developer implementations that are common in conventional parametric modeling tools.

\subsection{Agent Prompts}

Our system leverages structured prompting techniques \cite{white_prompt_2023} to guide LLMs toward producing responses in consistent and predictable formats.

To encourage active information gathering, we apply flipped interaction prompting \cite{white_prompt_2023} to the $R_A$, enabling it to proactively ask clarifying questions and autonomously guide the parametric modeling process. In contrast, the $C_A$ and $O_A$ operate in more linear roles: the $C_A$ handles code generation based on structured input, while the $O_A$ functions as an inspector, refining and validating the generated code for syntax and logic errors.

A key objective in designing these agent prompts was to ensure that each agent receives sufficient context while minimizing ambiguity. To this end, we applied XML tagging technique, which explicitly depict different information fragments, ensuring clear parsing and interpretation by each agent (see \cref{fig:prompt} for full structure).

\subsubsection{Reasoning Agent}

The Reasoning Agent serves as a domain expert in computational and parametric design for conceptual architectural tasks. Its primary role is to guide users in developing complete and well-structured parametric design specifications through structured dialogue. The RA prompt is composed of the following sections:
\begin{itemize}
    \item $<role>$ defines the RA’s identity and domain expertise.
    \item $<approach>$ specifies guidelines for conversational style and task objectives.
    \item $<parameters>$ describes the taxonomy of built-in and pre-defined data types (e.g., numeric values, geometric primitives, complex shapes) that constrain the agent’s outputs to implementable elements.
    \item $<user>$ defines the expected format for user inputs, which are wrapped in a standardized XML structure using <text> and <updates> tags.
    \item $<response>$ enforces a strict JSON schema ("type", "text", and "parameters") to ensure programmatic parsing and downstream compatibility.
    \item $<final>$ outlines the condition for concluding the interaction and initiating code generation.
\end{itemize}

$R_A$ progressively collects parameter specifications through multi-turn conversation, maintaining reference tracking until all required values are confirmed. Once complete, $R_A$ transitions to producing a structured instruction format for the downstream agents.

\subsubsection{Coding Agent \& Optimization Agent}

$C_A$ outputs structured C\# code blocks embedded with metadata, which are passed along for quality inspection by $O_A$. The $O_A$ conducts logic validation and structural optimization, outputting reviewed code blocks in a well-defined format. It supports a final exchange protocol between $C_A$ and $O_A$, where outputs include Name, Description, Dependency, and method-level breakdowns under Method\_New and Logic fields.

\subsection{Compiler and Execution}

To enable dynamic code execution within the Meta Quest 3’s constrained runtime environment, we introduce a two-device compiler architecture that enforces a clear separation between reusable methods and task-specific logic.

In this architecture, the Quest 3 acts as the server, while a networked desktop machine functions as a remote compilation client. This inversion ensures that the Quest can continue to operate stably even during network disruptions. A structured prompting protocol for a dedicated Coding Agent (See \cref{fig:prompt}) separates function definitions from procedural logic. Logic codes are only called to refresh visuals when parameter updates.

All user-defined functions are first extracted with explicit type signatures and compiled into a reusable DLL registry. Once all referenced methods are validated, the system compiles the high-level task logic, which orchestrates method calls through a unified interface. To minimize compile-time errors and ensure correctness, an Optimization Agent performs static analysis and dependency checks before execution. This inspection step guarantees that only fully resolvable code is deployed to the device.

The resulting system supports safe, flexible, and iterative authoring of parametric modeling code directly within immersive environments, despite the execution and compilation constraints from the hardware platform.


\section{Experiment and Discussion}

\subsection{Experiment Setup}

We implemented a two-phase experimental protocol for collecting feedback and acquiring both qualitative and quantitative observational data.

\textbf{Phase 1} involved a user study with 27 participants, including 10 design students and 17 professional practitioners. Participants interacted with our system to complete two predefined parametric design tasks  selected by the researchers, reflecting common design tasks in architectural design practice. Before start, participants completed a short questionnaire to assess prior knowledge of parametric design. Each interactive session lasted 15–30 minutes and was followed by a semi-structured interview to elicit subjective feedback on usability, interaction quality, and system intuitiveness. The total session time per participant was limited to 45 minutes.

During this phase, we systematically recorded dialogues between participants and RA, interaction durations, declared design intentions, and the system-generated parametric design outcomes irrespective of correctness. These multimodal data points formed the empirical basis for Phase 2.

\textbf{Phase 2} focused on evaluating the system’s reasoning-to-code mapping within an MR context. Using the parametric design task dataset collected in Phase 1 as input, researchers iteratively processed these through both the Coding Agent and Output Agent workflows. Our goal was to assess whether instructions targeting the MR context and robust code generation improved the semantic coherence and design fidelity of generated outputs. The resulting code and visual design artifacts were archived and independently evaluated by both architectural design experts and a subset of returning participants from Phase 1.

\subsection{User Study}

\subsubsection{Result}

\begin{figure}[tb]
 \centering 
 \includegraphics[width=\columnwidth]{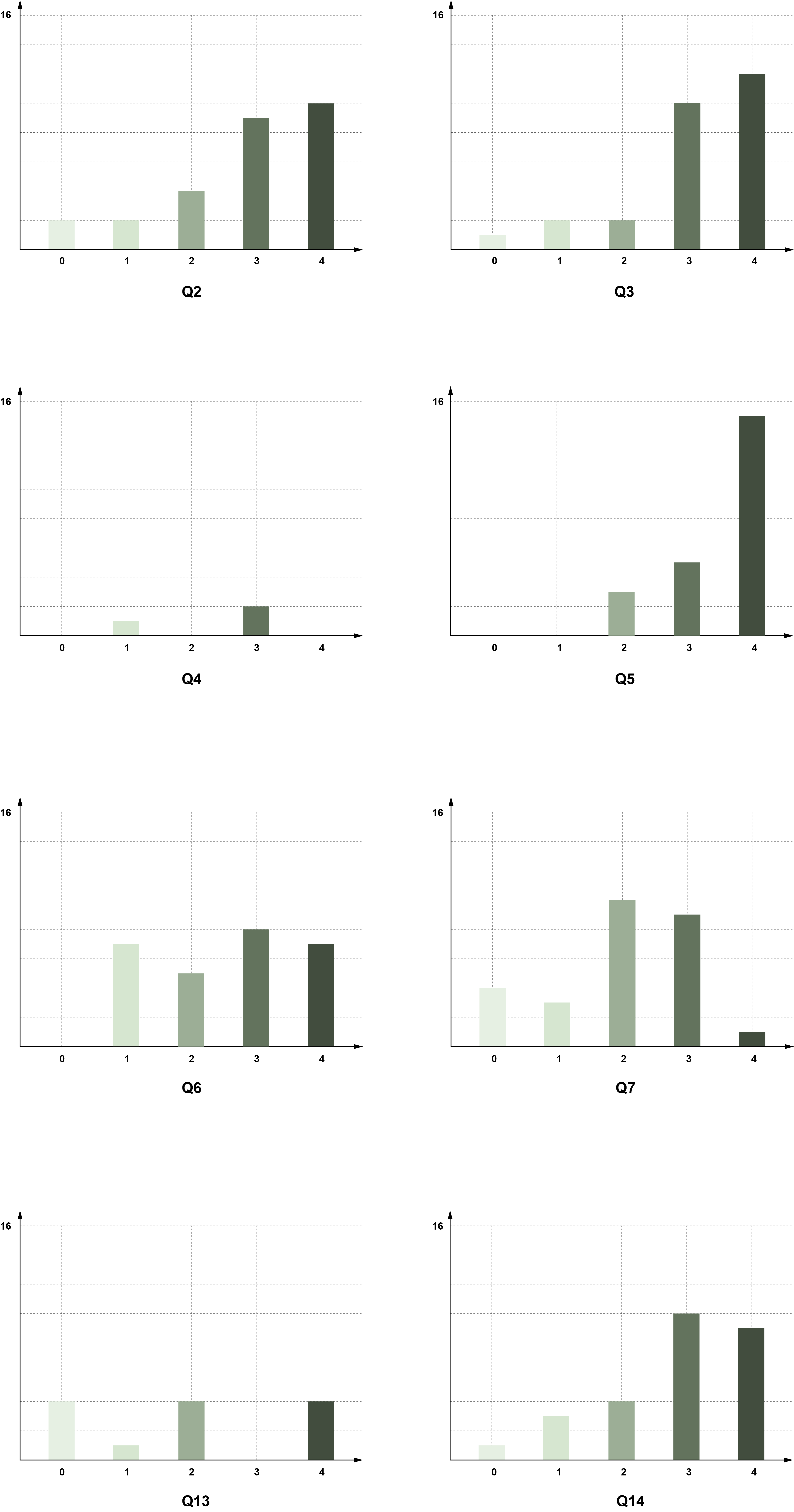}
 \caption{Result of user study questionnaire}
 \label{fig:q}
\end{figure}

\begin{figure}[tb]
 \centering 
 \includegraphics[width=\columnwidth]{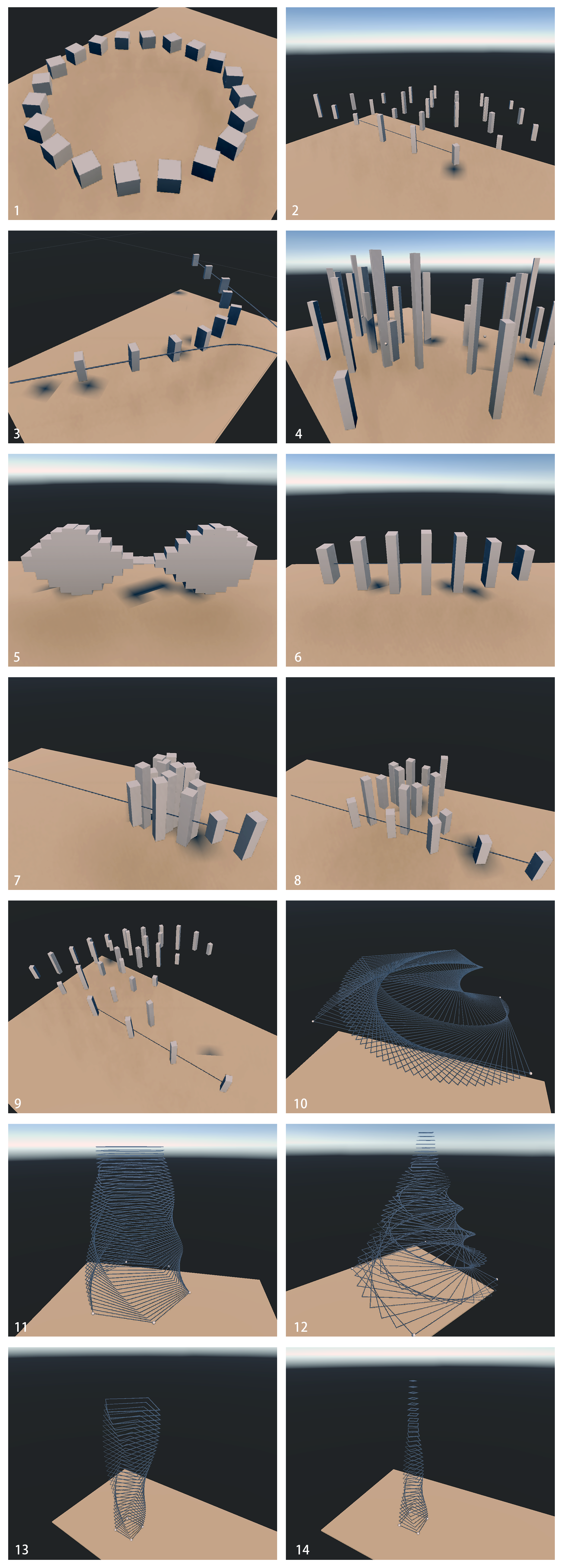}
 \caption{Result of parametric design generation result}
 \label{fig:results}
\end{figure}

In our user experiments, we selected two categories of tasks: Skyscraper generation and Planar generation (specifically, lot generation). These two tasks represent typical design requirements in architectural conceptual design. The representative design outcomes recorded during the experimental process are shown in the figure.

The 27 participants involved in the user study all came from architecture-related fields, including 10 architecture students and 17 architectural practitioners. According to the questionnaire results, at least 24 of the 27 participants either clearly understood the meaning of parametric design or had previously completed parametric design tasks, and they acknowledged the advantages of parametric design in improving design efficiency. Based on their feedback, Grasshopper was the most frequently mentioned parametric design software, greatly simplifying parametric design tasks with its visual programming approach. However, despite most participants being familiar with parametric design, 20 participants reported struggling with algorithmic implementation when faced with specific parametric design problems. Programming ability remains a primary obstacle limiting the broader application of parametric design within the architectural field. Our work is inspired by this challenge, aiming to explore possible solutions within an MR environment.

\subsubsection{Discussion}

By observing the process of participants performing parametric design tasks, we found that those who were more familiar with parametric design were able to quickly determine the specific parameters and corresponding values required to complete the design tasks through our conversational MR interface. In contrast, participants less familiar with parametric design were also able to gradually clarify the logical steps needed to complete the task and identify the necessary parameters and values through conversations with the large language model. Once the participants’ design intentions were understood and all required parameters and values were collected, our code generation system was able to stably output code that aligned with their design intentions and build the corresponding parametric model in the MR environment. This finding was also supported by participants’ feedback after the experiment: one participant mentioned, ``... it (RA) guided me through the parametric design process, and through the conversation with it, I gradually figured out the path that was previously unclear...”

\begin{figure}[tb]
 \centering 
 \includegraphics[width=\columnwidth]{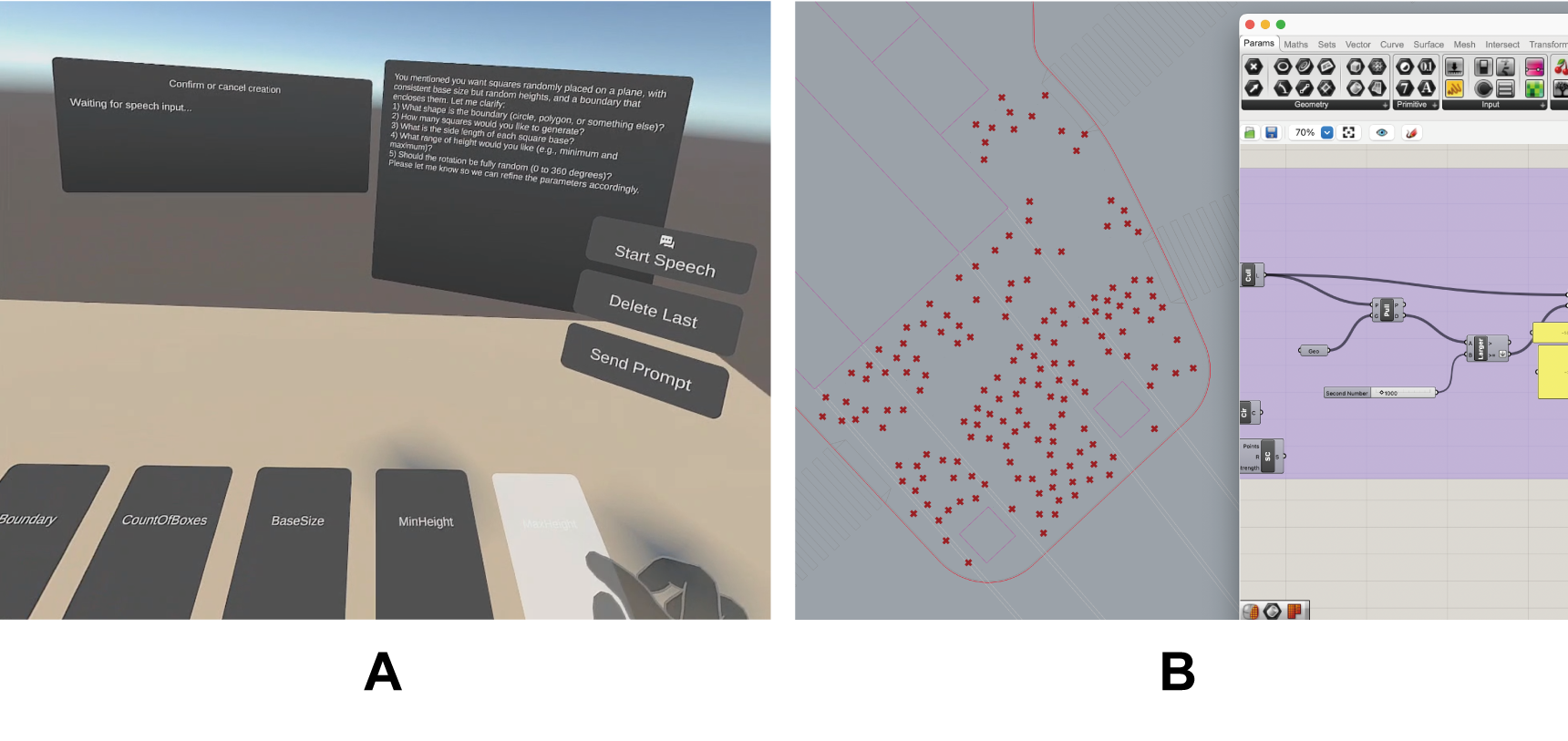}
 \caption{Comparison between A: Our proposed UI, and B: WIMP interface of Grasshopper in Rhino.}
 \label{fig:comparison}
\end{figure}

Most participants showed great interest in the generated parametric models, which can be attributed to the immersive interaction experience that only MR environments can provide, while comparing with existing tools like Grasshopper in \cref{fig:comparison}. They actively interacted with the models using the gestures we had presented, and some participants even explored further possibilities by changing parameters or logic after the interaction. Feedback also indicated that the effects of parameters on the models were more intuitively presented in the MR environment, helping to inspire designers’ creativity and imagination. These findings all point to the advantages of discussing parametric design within MR environments and further demonstrate the necessity of designing a parametric design interface that better fits MR interaction.


\subsection{Code Generation Experiment}

\subsubsection{Result}

In our code generation experiment, we aimed to temporarily set aside the goal of maintaining stable and error-free interaction processes, in order to investigate the conversion process from natural language to code in a multi-agent system within the context of parametric design tasks. Additionally, we sought to explore whether the specific content of instructions given to the $C_A$ would influence the outcomes of generated code.

Upon removing the $O_A$ from the system, compilation success rates declined significantly, dropping from 72.0\% (31/43) during user experiments to 25.7\% in our testing environment (9/35).

\subsubsection{Discussion}

Our data generation experiments further discover the nature of the proposed ``reasoning-to-code-generation” pipeline operating within MR environments. During the experiments, we observed an intriguing phenomenon: our designed system exhibits a trade-off between accuracy and creativity.

Specifically, we found that incorporating MR-specific prompts or instructions that precisely define tasks—initially designed by researchers to prevent $C_A$ from generating code that conflicts with common sense in architectural design—led to more stable and reliably executable code. In contrast, when these contextual prompts were omitted, the $C_A$ generated code using more diverse and inventive approaches, though often at the cost of syntactic or semantic correctness.

For example, we compared the effects of adding specific instructions in $R_A$'s $<final>$ tag, which determines the instructions given to the $C_A$. The description in $<final>$ changed dramatically in our comparisons. While this change aligns with the capabilities of modern LLMs, we want to highlight how it affected $C_A$'s generation results for the same task. Specifically, prompts with strict definitions produced highly similar results across generations, while more flexible prompts produced code outputs with consistent main logic but varying numbers of new methods defined by the LLM.

\begin{itemize}
    \item Added prompt: ``It should clearly reference how to use parameters like a variable in each step precisely.""
    \item Before: ``This planar distribution logic creates two adjacent squares on a horizontal plane, each defined as an S2MClosedPolyline parameter. The first square's vertices span from (0,0,0) to (1,1,0), while the second square is offset horizontally by 2 units..."

    \item After: ``...design logic arranges concentric ovals in plan around a common center [BaseCenter]. The user sets [OvalCount] to define how many ovals to create. The ovals grow outward from an initial size defined by [InitialMajorRadius] and [InitialMinorRadius]..."
\end{itemize}

This observation suggests that while stability and executability are crucial for maintaining system reliability, encouraging variability in code generation holds significant potential for stimulating creative exploration—an essential aspect of early-stage architectural design. As also reflected in our user study, conversational MR interfaces can help guide and inspire participants’ design thinking. Similarly, we argue that preserving creative latitude within the code generation process can enhance users’ sense of authorship and engagement.

Balancing this trade-off between correctness and creativity will be a primary focus in our future optimization efforts. We envision approaches such as dynamic prompt tuning, adaptive reasoning strategies, or tiered validation layers as potential pathways to maintain code stability while fostering exploratory generation.

\subsubsection{Conclusion}

Our evaluation indicates that the system possesses foundational capabilities to support parametric modeling within an MR environment. Users were able to complete modeling tasks with relatively few compilation failures under the set of $C_A$ and $O_A$, and the interactive framework contributed to improved understanding of design logic and creative ideation. These results suggest that the integration of conversational interfaces and MR can lower the barrier to entry for parametric design, particularly for users without prior coding experience.

However, the study also uncovered areas requiring further investigation. While we observed instances where LLMs exhibited a rudimentary ability to interpret spatial relationships, we did not conduct a systematic comparison of reasoning models or formally assess spatial awareness. As MR-based design tasks often involve complex 3D spatial reasoning, future work should focus on evaluating and enhancing the spatial understanding of LLMs.

In addition, improvements in prompt engineering—particularly those that tailor inputs for spatially grounded tasks—could enhance both the reliability and relevance of code generation. We also identify multi-agent collaboration, especially through token-level communication protocols, as a promising direction for future research. Such mechanisms may enable more sophisticated reasoning, improved error handling, and collaborative problem-solving within the MR design process.


\section{Conclusion and Future Development}

In this paper, we have presented methods that integrate parametric design tools into MR environments for architectural concept design through a conversational interface driven by MR and LLM-agent-based interaction. Through user studies and code generation experiments based on user study data, we evaluated the system’s reliability, user feedback, and interaction performance. The results demonstrate that the actual user interaction goals and workflows were generally consistent with our design expectations. The ``reasoning-to-code-generation approach” effectively reduces the barrier for designers with limited programming or algorithmic experience, enabling them to focus more on the logic and spatial relationships of parametric design rather than the construction of algorithms.

Beyond the initial research questions, user feedback revealed the unique advantages of combining conversational MR interfaces, especially in guiding users through design exploration and helping them gradually clarify their design intentions. Unlike traditional WIMP interfaces where users approach the system with predefined goals and fixed set of functions, our approach encourages an open-ended, conversational exploration of possibilities until a satisfactory outcome is achieved. Furthermore, during the evaluation of generated results—both successful and unsuccessful—we encountered preliminary observations regarding the spatial awareness capabilities of LLMs, a topic that has received increasing attention in related research. We believe similar challenges, such as those addressed in embodied intelligence and decision-making, will emerge when applying LLMs in MR contexts.

Despite the promising outcomes, this study also has several limitations that suggest directions for future development. Firstly, in terms of interaction, the current system supports only basic gesture and language inputs. Future iterations could explore richer multimodal inputs, such as gaze tracking and voice emphasis to enhance spatial referencing and intuitive model manipulation in MR. Secondly, in terms of parametric functionality, our system does not yet support more complex design parameters such as materials, colors, or free-form geometries. While developing a complete and robust shape classification system lies beyond the scope of this paper, it is a critical step for future architectural design applications. Moreover, algorithmic design capabilities such as optimization, simulation, and performance analysis have not been integrated, though the modularity and flexibility of the current system make such extensions feasible and promising.

In conclusion, we believe the MR+LLM parametric design interaction framework presented in this study offers a novel direction for architectural parametric design. It not only improves design efficiency and stimulates creativity but also lays a solid foundation for the future evolution of intelligent design systems in architecture.



\bibliographystyle{abbrv-doi}

\bibliography{zotero_s2m}

\begin{thebibliography}{10}

\bibitem{noauthor_httpswwwgrasshopper3dcom_nodate}
https://www.grasshopper3d.com/.

\bibitem{ahn_as_2022}
M.~Ahn, A.~Brohan, N.~Brown, Y.~Chebotar, O.~Cortes, B.~David, C.~Finn, C.~Fu, K.~Gopalakrishnan, K.~Hausman, A.~Herzog, D.~Ho, J.~Hsu, J.~Ibarz, B.~Ichter, A.~Irpan, E.~Jang, R.~J. Ruano, K.~Jeffrey, S.~Jesmonth, N.~J. Joshi, R.~Julian, D.~Kalashnikov, Y.~Kuang, K.-H. Lee, S.~Levine, Y.~Lu, L.~Luu, C.~Parada, P.~Pastor, J.~Quiambao, K.~Rao, J.~Rettinghouse, D.~Reyes, P.~Sermanet, N.~Sievers, C.~Tan, A.~Toshev, V.~Vanhoucke, F.~Xia, T.~Xiao, P.~Xu, S.~Xu, M.~Yan, and A.~Zeng.
\newblock Do as i can, not as i say: Grounding language in robotic affordances.
\newblock Version Number: 2. doi: {{%
10\hspace{.1pt}\discretionary{.}{%
}{.}\hspace{.4pt}48550\discretionary{/}{%
}{/}ARXIV\hspace{.1pt}\discretionary{.}{%
}{.}\hspace{.4pt}2204\hspace{.1pt}\discretionary{.}{%
}{.}\hspace{.4pt}01691}}


\bibitem{alhazzaa_integrating_2023}
K.~Alhazzaa and W.~Yan.
\newblock Integrating parametric modeling, {BIM}, and building performance analysis into augmented reality for architectural design and education.
\newblock In {\em Proceedings of the 2023 7th International Conference on Virtual and Augmented Reality Simulations}, pp. 68--76. {ACM}. doi: {{%
10\hspace{.1pt}\discretionary{.}{%
}{.}\hspace{.4pt}1145\discretionary{/}{%
}{/}3603421\hspace{.1pt}\discretionary{.}{%
}{.}\hspace{.4pt}3603431}}


\bibitem{alibay_usability_2017}
F.~Alibay, M.~Kavakli, J.-R. Chardonnet, and M.~Z. Baig.
\newblock The usability of speech and/or gestures in multi-modal interface systems.
\newblock In {\em Proceedings of the 9th International Conference on Computer and Automation Engineering}, pp. 73--77. {ACM}. doi: {{%
10\hspace{.1pt}\discretionary{.}{%
}{.}\hspace{.4pt}1145\discretionary{/}{%
}{/}3057039\hspace{.1pt}\discretionary{.}{%
}{.}\hspace{.4pt}3057089}}


\bibitem{alrashedy_generating_2024}
K.~Alrashedy, P.~Tambwekar, Z.~Zaidi, M.~Langwasser, W.~Xu, and M.~Gombolay.
\newblock Generating {CAD} code with vision-language models for 3d designs.
\newblock Version Number: 1. doi: {{%
10\hspace{.1pt}\discretionary{.}{%
}{.}\hspace{.4pt}48550\discretionary{/}{%
}{/}ARXIV\hspace{.1pt}\discretionary{.}{%
}{.}\hspace{.4pt}2410\hspace{.1pt}\discretionary{.}{%
}{.}\hspace{.4pt}05340}}


\bibitem{baig_analyzing_2018}
M.~Baig and M.~Kavakli.
\newblock Analyzing novice and expert user's cognitive load in using a multi-modal interface system. doi: {{%
10\hspace{.1pt}\discretionary{.}{%
}{.}\hspace{.4pt}1109\discretionary{/}{%
}{/}ICSENG\hspace{.1pt}\discretionary{.}{%
}{.}\hspace{.4pt}2018\hspace{.1pt}\discretionary{.}{%
}{.}\hspace{.4pt}8638206}}


\bibitem{bhooshan_parametric_2017}
S.~Bhooshan.
\newblock Parametric design thinking: A case-study of practice-embedded architectural research.
\newblock 52:115--143. doi: {{%
10\hspace{.1pt}\discretionary{.}{%
}{.}\hspace{.4pt}1016\discretionary{/}{%
}{/}j\hspace{.1pt}\discretionary{.}{%
}{.}\hspace{.4pt}destud\hspace{.1pt}\discretionary{.}{%
}{.}\hspace{.4pt}2017\hspace{.1pt}\discretionary{.}{%
}{.}\hspace{.4pt}05\hspace{.1pt}\discretionary{.}{%
}{.}\hspace{.4pt}003}}


\bibitem{bolt_put-that-there_1980}
R.~A. Bolt.
\newblock “put-that-there”: Voice and gesture at the graphics interface.
\newblock 14(3):262--270. doi: {{%
10\hspace{.1pt}\discretionary{.}{%
}{.}\hspace{.4pt}1145\discretionary{/}{%
}{/}965105\hspace{.1pt}\discretionary{.}{%
}{.}\hspace{.4pt}807503}}


\bibitem{buyruk_interactive_2022}
Y.~Buyruk and G.~Çağdaş.
\newblock Interactive parametric design and robotic fabrication within mixed reality environment.
\newblock 12(24):12797. doi: {{%
10\hspace{.1pt}\discretionary{.}{%
}{.}\hspace{.4pt}3390\discretionary{/}{%
}{/}app122412797}}


\bibitem{caetano_computational_2020}
I.~Caetano, L.~Santos, and A.~Leitão.
\newblock Computational design in architecture: Defining parametric, generative, and algorithmic design.
\newblock 9(2):287--300. doi: {{%
10\hspace{.1pt}\discretionary{.}{%
}{.}\hspace{.4pt}1016\discretionary{/}{%
}{/}j\hspace{.1pt}\discretionary{.}{%
}{.}\hspace{.4pt}foar\hspace{.1pt}\discretionary{.}{%
}{.}\hspace{.4pt}2019\hspace{.1pt}\discretionary{.}{%
}{.}\hspace{.4pt}12\hspace{.1pt}\discretionary{.}{%
}{.}\hspace{.4pt}008}}


\bibitem{cai_3description_2024}
Z.~Cai.
\newblock 3description: An intuitive human-{AI} collaborative 3d modeling approach.
\newblock In {\em Proceedings of the 11th International Conference on Digital and Interactive Arts}, {ARTECH} '23. Association for Computing Machinery.
\newblock event-place: Faro, Portugal. doi: {{%
10\hspace{.1pt}\discretionary{.}{%
}{.}\hspace{.4pt}1145\discretionary{/}{%
}{/}3632776\hspace{.1pt}\discretionary{.}{%
}{.}\hspace{.4pt}3632785}}


\bibitem{castelo-branco_algorithmic_2022}
R.~Castelo-Branco and A.~Leitão.
\newblock Algorithmic design in virtual reality.
\newblock 2(1):31--52. doi: {{%
10\hspace{.1pt}\discretionary{.}{%
}{.}\hspace{.4pt}3390\discretionary{/}{%
}{/}architecture2010003}}


\bibitem{chatterjee_free-form_2024}
S.~Chatterjee.
\newblock Free-form shape modeling in {XR}: A systematic review.
\newblock Version Number: 1. doi: {{%
10\hspace{.1pt}\discretionary{.}{%
}{.}\hspace{.4pt}48550\discretionary{/}{%
}{/}ARXIV\hspace{.1pt}\discretionary{.}{%
}{.}\hspace{.4pt}2401\hspace{.1pt}\discretionary{.}{%
}{.}\hspace{.4pt}00924}}


\bibitem{cheng_empowering_2024}
G.~Cheng, C.~Zhang, W.~Cai, L.~Zhao, C.~Sun, and J.~Bian.
\newblock Empowering large language models on robotic manipulation with affordance prompting.
\newblock Version Number: 1. doi: {{%
10\hspace{.1pt}\discretionary{.}{%
}{.}\hspace{.4pt}48550\discretionary{/}{%
}{/}ARXIV\hspace{.1pt}\discretionary{.}{%
}{.}\hspace{.4pt}2404\hspace{.1pt}\discretionary{.}{%
}{.}\hspace{.4pt}11027}}


\bibitem{coppens_parametric_2018}
A.~Coppens, T.~Mens, and M.-A. Gallas.
\newblock Parametric modelling within immersive environments - building a bridge between existing tools and virtual reality headsets.
\newblock pp. 711--716. doi: {{%
10\hspace{.1pt}\discretionary{.}{%
}{.}\hspace{.4pt}52842\discretionary{/}{%
}{/}conf\hspace{.1pt}\discretionary{.}{%
}{.}\hspace{.4pt}ecaade\hspace{.1pt}\discretionary{.}{%
}{.}\hspace{.4pt}2018\hspace{.1pt}\discretionary{.}{%
}{.}\hspace{.4pt}2\hspace{.1pt}\discretionary{.}{%
}{.}\hspace{.4pt}711}}


\bibitem{desolda_digital_2023}
G.~Desolda, A.~Esposito, F.~Müller, and S.~Feger.
\newblock Digital modeling for everyone: Exploring how novices approach voice-based 3d modeling.
\newblock vol. 14145 {LNCS}, pp. 133--155. doi: {{%
10\hspace{.1pt}\discretionary{.}{%
}{.}\hspace{.4pt}1007\discretionary{/}{%
}{/}978\discretionary{%
}{-}{-}3\discretionary{%
}{-}{-}031\discretionary{%
}{-}{-}42293\discretionary{%
}{-}{-}5\_11}}


\bibitem{fang_comparisons_2023}
Y.-M. Fang and T.-L. Kao.
\newblock Comparisons of emotional responses, flow experiences, and operational performances in traditional parametric computer-aided design modeling and virtual-reality free-form modeling.
\newblock 13(11):6568. doi: {{%
10\hspace{.1pt}\discretionary{.}{%
}{.}\hspace{.4pt}3390\discretionary{/}{%
}{/}app13116568}}


\bibitem{friedrich_combining_2021}
M.~Friedrich, S.~Langer, and F.~Frey.
\newblock Combining gesture and voice control for mid-air manipulation of {CAD} models in {VR} environments.
\newblock vol.~2, pp. 119--127.

\bibitem{gao_pal_2022}
L.~Gao, A.~Madaan, S.~Zhou, U.~Alon, P.~Liu, Y.~Yang, J.~Callan, and G.~Neubig.
\newblock {PAL}: Program-aided language models.
\newblock Version Number: 2. doi: {{%
10\hspace{.1pt}\discretionary{.}{%
}{.}\hspace{.4pt}48550\discretionary{/}{%
}{/}ARXIV\hspace{.1pt}\discretionary{.}{%
}{.}\hspace{.4pt}2211\hspace{.1pt}\discretionary{.}{%
}{.}\hspace{.4pt}10435}}


\bibitem{giunchi_mixing_2021}
D.~Giunchi, A.~Sztrajman, S.~James, and A.~Steed.
\newblock Mixing modalities of 3d sketching and speech for interactive model retrieval in virtual reality.
\newblock In {\em Proceedings of the 2021 {ACM} International Conference on Interactive Media Experiences}, {IMX} '21, pp. 144--155. Association for Computing Machinery.
\newblock event-place: Virtual Event, {USA}. doi: {{%
10\hspace{.1pt}\discretionary{.}{%
}{.}\hspace{.4pt}1145\discretionary{/}{%
}{/}3452918\hspace{.1pt}\discretionary{.}{%
}{.}\hspace{.4pt}3458806}}


\bibitem{gurel_cognitive_2023}
A.~Gürel and B.~Şenyapılı Ozcan.
\newblock Cognitive comparison of design methods in the conceptual phase.
\newblock 21(4):581--601. doi: {{%
10\hspace{.1pt}\discretionary{.}{%
}{.}\hspace{.4pt}1177\discretionary{/}{%
}{/}14780771231188474}}


\bibitem{hong_3d-llm_2023}
Y.~Hong, H.~Zhen, P.~Chen, S.~Zheng, Y.~Du, Z.~Chen, and C.~Gan.
\newblock 3d-{LLM}: Injecting the 3d world into large language models.
\newblock Version Number: 1. doi: {{%
10\hspace{.1pt}\discretionary{.}{%
}{.}\hspace{.4pt}48550\discretionary{/}{%
}{/}ARXIV\hspace{.1pt}\discretionary{.}{%
}{.}\hspace{.4pt}2307\hspace{.1pt}\discretionary{.}{%
}{.}\hspace{.4pt}12981}}


\bibitem{huang_instruct2act_2023}
S.~Huang, Z.~Jiang, H.~Dong, Y.~Qiao, P.~Gao, and H.~Li.
\newblock Instruct2act: Mapping multi-modality instructions to robotic actions with large language model.
\newblock Version Number: 3. doi: {{%
10\hspace{.1pt}\discretionary{.}{%
}{.}\hspace{.4pt}48550\discretionary{/}{%
}{/}ARXIV\hspace{.1pt}\discretionary{.}{%
}{.}\hspace{.4pt}2305\hspace{.1pt}\discretionary{.}{%
}{.}\hspace{.4pt}11176}}


\bibitem{hollein_text2room_2023}
L.~Höllein, A.~Cao, A.~Owens, J.~Johnson, and M.~Nießner.
\newblock Text2room: Extracting textured 3d meshes from 2d text-to-image models.
\newblock Version Number: 2. doi: {{%
10\hspace{.1pt}\discretionary{.}{%
}{.}\hspace{.4pt}48550\discretionary{/}{%
}{/}ARXIV\hspace{.1pt}\discretionary{.}{%
}{.}\hspace{.4pt}2303\hspace{.1pt}\discretionary{.}{%
}{.}\hspace{.4pt}11989}}


\bibitem{lee_creativity_2015}
J.~H. Lee, N.~Gu, and M.~J. Ostwald.
\newblock Creativity and parametric design? comparing designer's cognitive approaches with assessed levels of creativity.
\newblock 3(2):78--94. doi: {{%
10\hspace{.1pt}\discretionary{.}{%
}{.}\hspace{.4pt}1080\discretionary{/}{%
}{/}21650349\hspace{.1pt}\discretionary{.}{%
}{.}\hspace{.4pt}2014\hspace{.1pt}\discretionary{.}{%
}{.}\hspace{.4pt}931826}}


\bibitem{liang_code_2023}
J.~Liang, W.~Huang, F.~Xia, P.~Xu, K.~Hausman, B.~Ichter, P.~Florence, and A.~Zeng.
\newblock Code as policies: Language model programs for embodied control.
\newblock In {\em 2023 {IEEE} International Conference on Robotics and Automation ({ICRA})}, pp. 9493--9500. {IEEE}. doi: {{%
10\hspace{.1pt}\discretionary{.}{%
}{.}\hspace{.4pt}1109\discretionary{/}{%
}{/}ICRA48891\hspace{.1pt}\discretionary{.}{%
}{.}\hspace{.4pt}2023\hspace{.1pt}\discretionary{.}{%
}{.}\hspace{.4pt}10160591}}


\bibitem{lin_text2motion_2023}
K.~Lin, C.~Agia, T.~Migimatsu, M.~Pavone, and J.~Bohg.
\newblock Text2motion: from natural language instructions to feasible plans.
\newblock 47(8):1345--1365. doi: {{%
10\hspace{.1pt}\discretionary{.}{%
}{.}\hspace{.4pt}1007\discretionary{/}{%
}{/}s10514\discretionary{%
}{-}{-}023\discretionary{%
}{-}{-}10131\discretionary{%
}{-}{-}7}}


\bibitem{liu_llmp_2023}
B.~Liu, Y.~Jiang, X.~Zhang, Q.~Liu, S.~Zhang, J.~Biswas, and P.~Stone.
\newblock {LLM}+p: Empowering large language models with optimal planning proficiency.
\newblock Version Number: 3. doi: {{%
10\hspace{.1pt}\discretionary{.}{%
}{.}\hspace{.4pt}48550\discretionary{/}{%
}{/}ARXIV\hspace{.1pt}\discretionary{.}{%
}{.}\hspace{.4pt}2304\hspace{.1pt}\discretionary{.}{%
}{.}\hspace{.4pt}11477}}


\bibitem{salim_system_2010}
F.~Salim, H.~Mulder, and J.~Burry.
\newblock A system for form fostering: Parametric modeling of responsive forms in mixed reality.
\newblock pp. 531--540. doi: {{%
10\hspace{.1pt}\discretionary{.}{%
}{.}\hspace{.4pt}52842\discretionary{/}{%
}{/}conf\hspace{.1pt}\discretionary{.}{%
}{.}\hspace{.4pt}caadria\hspace{.1pt}\discretionary{.}{%
}{.}\hspace{.4pt}2010\hspace{.1pt}\discretionary{.}{%
}{.}\hspace{.4pt}531}}


\bibitem{sun_3d-gpt_2023}
C.~Sun, J.~Han, W.~Deng, X.~Wang, Z.~Qin, and S.~Gould.
\newblock 3d-{GPT}: Procedural 3d modeling with large language models.
\newblock Version Number: 2. doi: {{%
10\hspace{.1pt}\discretionary{.}{%
}{.}\hspace{.4pt}48550\discretionary{/}{%
}{/}ARXIV\hspace{.1pt}\discretionary{.}{%
}{.}\hspace{.4pt}2310\hspace{.1pt}\discretionary{.}{%
}{.}\hspace{.4pt}12945}}


\bibitem{vemprala_chatgpt_2024}
S.~H. Vemprala, R.~Bonatti, A.~Bucker, and A.~Kapoor.
\newblock {ChatGPT} for robotics: Design principles and model abilities.
\newblock 12:55682--55696. doi: {{%
10\hspace{.1pt}\discretionary{.}{%
}{.}\hspace{.4pt}1109\discretionary{/}{%
}{/}ACCESS\hspace{.1pt}\discretionary{.}{%
}{.}\hspace{.4pt}2024\hspace{.1pt}\discretionary{.}{%
}{.}\hspace{.4pt}3387941}}


\bibitem{wang_describe_2023}
Z.~Wang, S.~Cai, G.~Chen, A.~Liu, X.~Ma, and Y.~Liang.
\newblock Describe, explain, plan and select: Interactive planning with large language models enables open-world multi-task agents.
\newblock Version Number: 3. doi: {{%
10\hspace{.1pt}\discretionary{.}{%
}{.}\hspace{.4pt}48550\discretionary{/}{%
}{/}ARXIV\hspace{.1pt}\discretionary{.}{%
}{.}\hspace{.4pt}2302\hspace{.1pt}\discretionary{.}{%
}{.}\hspace{.4pt}01560}}


\bibitem{weng_dream_2024}
S.-C. Weng, Y.-M. Chiou, and E.-L. Do.
\newblock Dream mesh: A speech-to-3d model generative pipeline in mixed reality.
\newblock pp. 345--349. doi: {{%
10\hspace{.1pt}\discretionary{.}{%
}{.}\hspace{.4pt}1109\discretionary{/}{%
}{/}AIxVR59861\hspace{.1pt}\discretionary{.}{%
}{.}\hspace{.4pt}2024\hspace{.1pt}\discretionary{.}{%
}{.}\hspace{.4pt}00059}}


\bibitem{white_prompt_2023}
J.~White, Q.~Fu, S.~Hays, M.~Sandborn, C.~Olea, H.~Gilbert, A.~Elnashar, J.~Spencer-Smith, and D.~C. Schmidt.
\newblock A prompt pattern catalog to enhance prompt engineering with {ChatGPT}. doi: {{%
10\hspace{.1pt}\discretionary{.}{%
}{.}\hspace{.4pt}48550\discretionary{/}{%
}{/}arXiv\hspace{.1pt}\discretionary{.}{%
}{.}\hspace{.4pt}2302\hspace{.1pt}\discretionary{.}{%
}{.}\hspace{.4pt}11382}}


\bibitem{williams_understanding_2020}
A.~Williams, J.~Garcia, and F.~Ortega.
\newblock Understanding multimodal user gesture and speech behavior for object manipulation in augmented reality using elicitation.
\newblock 26(12):3479--3489. doi: {{%
10\hspace{.1pt}\discretionary{.}{%
}{.}\hspace{.4pt}1109\discretionary{/}{%
}{/}TVCG\hspace{.1pt}\discretionary{.}{%
}{.}\hspace{.4pt}2020\hspace{.1pt}\discretionary{.}{%
}{.}\hspace{.4pt}3023566}}


\bibitem{wortmann_differentiating_2017}
T.~Wortmann and B.~Tunçer.
\newblock Differentiating parametric design: Digital workflows in contemporary architecture and construction.
\newblock 52:173--197. doi: {{%
10\hspace{.1pt}\discretionary{.}{%
}{.}\hspace{.4pt}1016\discretionary{/}{%
}{/}j\hspace{.1pt}\discretionary{.}{%
}{.}\hspace{.4pt}destud\hspace{.1pt}\discretionary{.}{%
}{.}\hspace{.4pt}2017\hspace{.1pt}\discretionary{.}{%
}{.}\hspace{.4pt}05\hspace{.1pt}\discretionary{.}{%
}{.}\hspace{.4pt}004}}


\bibitem{yin_text2vrscene_2024}
Z.~Yin, Y.~Wang, T.~Papatheodorou, and P.~Hui.
\newblock Text2vrscene: Exploring the framework of automated text-driven generation system for {VR} experience.
\newblock In {\em 2024 {IEEE} Conference Virtual Reality and 3D User Interfaces ({VR})}, pp. 701--711. doi: {{%
10\hspace{.1pt}\discretionary{.}{%
}{.}\hspace{.4pt}1109\discretionary{/}{%
}{/}VR58804\hspace{.1pt}\discretionary{.}{%
}{.}\hspace{.4pt}2024\hspace{.1pt}\discretionary{.}{%
}{.}\hspace{.4pt}00090}}


\bibitem{zhou_eliciting_2022}
X.~Zhou, A.~Williams, and F.~Ortega.
\newblock Eliciting multimodal gesture+speech interactions in a multi-object augmented reality environment. doi: {{%
10\hspace{.1pt}\discretionary{.}{%
}{.}\hspace{.4pt}1145\discretionary{/}{%
}{/}3562939\hspace{.1pt}\discretionary{.}{%
}{.}\hspace{.4pt}3565637}}


\end{thebibliography}

\end{document}